\newcommand{\al} {\alpha}
\newcommand{\bl} {\beta}
\newcommand{\cl} {\gamma}
\newcommand{\dl} {\delta}

\newcommand{\bee} {\begin{equation}}
\newcommand{\ene} {\end{equation}}
\newcommand{\beqa} {\begin{eqnarray}}
\newcommand{\enqa} {\end{eqnarray}}
\newcommand{\beqsa} {\begin{eqnarray*}}
\newcommand{\enqsa} {\end{eqnarray*}}
\newcommand{\bea} {\begin{array}}
\newcommand{\ena} {\end{array}}

\newcommand{\gp} {\Phi}
\newcommand{\p} {\psi}
\newcommand{\rw} {\rightarrow}

\documentstyle[11pt]{article}
\begin{document}
\baselineskip=18pt
\noindent

\def\bt{\tilde\beta}
\def\gt{\tilde\gamma}
\def\dt{\tilde\delta}
\def\fut{\tilde f_1}
\def\fdt{\tilde f_2}
\def\ftt{\tilde f_3}
\def\t{\tau_{p\rightarrow e^+\pi^0}}
\def\aas{{\alpha\over\alpha_S}(M_Z)}
\def\t{\tau_{p\rw e^+\pi^0}}
\def\si{\sin^2{\theta_W(M_Z)}}
\def\sq{SU(4)_{PS}\otimes SU(2)_L\otimes SU(2)_R}
\def\stt{SU(3)_c\otimes SU(2)_L\otimes SU(2)_R\otimes U(1)_{B-L}}
\def\st{SU(3)_c\otimes SU(2)_L\otimes SU(2)_R}

{\hfill \bf DSF-93/52~~~~~~~~}

{\hfill \bf MIT-CTP-2286~~~~~~}

\vskip 1cm

\centerline{\bf PROTON DECAY AND NEUTRINO MASSES IN SO(10)}

\bigskip

\centerline{\begin{tabular}{lll}
F. Acampora$^*$, & G. Amelino-Camelia$^\triangle$, &
F. Buccella$^{\dagger *}$, \\
O. Pisanti$^{\dagger\ddagger}$, & L. Rosa$^{\dagger\ddagger}$~~and
& T. Tuzi$^*$
\end{tabular}}

\begin{itemize}
\item[*] Istituto di Fisica Teorica, Mostra d'Oltremare, 80125 Napoli, Italia.
\item[$\triangle$] Center for Theoretical Physics, Laboratory of Nuclear
Science and Department of Physics, Massachusetts Institute of Technology,
Cambridge, Massachusetts 02139, USA.
\item[$\dagger$] Dipartimento di Scienze Fisiche, Mostra d'Oltremare
Pad. 19, 80125 Napoli, Italia.
\item[$\ddagger$] Istituto Nazionale di Fisica Nucleare, Mostra d'Oltremare
Pad. 20, 80125 Napoli, Italia.
\end{itemize}

\begin{abstract}
In the last few years physicists have been looking at $SO(10)$ GUT
models with renewed attention because it has been realized that the SU(5)
minimal model cannot unify the strong, electromagnetic and weak interactions
in a way consistent with the experimental values of $\alpha(M_Z)$,
$\si$ and $\alpha_S(M_Z)$. In this paper
we derive lower limits on neutrino masses, relevant for cosmology and for
the solar-neutrino problem, from necessary consistency conditions on a
class of $SO(10)$ models with $\sq$ or $\stt$ intermediate gauge symmetry.
\end{abstract}

\vspace{\bigskipamount}

{\footnotesize

CONTENTS

I. Introduction \hfill 2

II. Symmetry Breaking for Various Higgs Potentials \hfill 6

\indent \indent A. $G'\equiv SU(4)_{PS}\otimes SU(2)_L\otimes SU(2)_R\times
D$ \hfill 6

\indent \indent B. The 210 representation \hfill 8

\indent \indent \indent 1. $G'\equiv SU(4)_{PS}\otimes SU(2)_L\otimes SU(2)_R$
\hfill 14

\indent \indent \indent 2. $G'\equiv SU(3)_c\otimes SU(2)_L\otimes SU(2)_R
\otimes U(1)_{B-L}\times D$ \hfill 15

\indent \indent \indent 3. $G'\equiv SU(3)_c\otimes SU(2)_L\otimes SU(2)_R
\otimes U(1)_{B-L}$ \hfill 16

III. Renormalization-Group Equations \hfill 18

\indent \indent A. $G'\equiv SU(4)\otimes SU(2)\otimes SU(2)\times D$ \hfill 20

\indent \indent B. $G'\equiv SU(4)_c\otimes SU(2)_L\otimes SU(2)_R$ \hfill 21

\indent \indent C. $G'\equiv SU(3)_c\otimes SU(2)_L\otimes SU(2)_R\otimes
U(1)_{B-L}
\times D$ \hfill 23

\indent \indent D. $G'\equiv SU(3)_c\otimes SU(2)_L\otimes SU(2)_R\otimes
U(1)_{B-L}$
\hfill 24

IV. Conclusions \hfill 26

Appendix \hfill 28

\indent References  \hfill 31

}

\section*{I. Introduction}

A $SO(10)$ (Georgi, 1975; Fritzsch and Minkowski, 1975) unified model
with intermediate symmetry
$SU(4)_{PS}\otimes SU(2)_L\otimes SU(2)_R\equiv G_{422}$
(Pati and Salam, 1974; Chang, Mohapatra, and Parida, 1984;
Chang, Gipson, Marshak, Mohapatra, and Parida, 1985),
broken at a scale $M_R\sim 10^{11}~GeV$, has been
recently advocated (Mohapatra and Parida, 1993; Babu and Shafi, 1993)
to give, through the see-saw mechanism (Gell-Mann, Ramond, and Slansky, 1980;
Yanagida, 1979), neutrino masses of the order required to explain the
solar-neutrino problem within the framework of the MSW theory, and to account
for at least part of the dark matter in the universe (Schramm, 1992).

$SO(10)$ unified models have been studied since many years with the
physical motivation of obtaining values for the masses of the
lepto-quarks which mediate proton decay higher than the ones found within
the $SU(5)$ minimal model. The $SU(5)$ predictions for
these masses are too low with respect to the experimental lower limit on
proton decay: $\t\geq9\cdot10^{32}~years$ (Review of Particle Properties,
1992), which corresponds (le Yaouanc, Oliver, P\`ene, and Raynal, 1977)
to the following lower limit on the masses (which we indicate
with $M_X$) of the lepto-quarks which mediate that decay:
\bee
M_X\geq 3.2\cdot10^{15}~GeV. \label{eq:mxinf}
\ene

\begin{figure}[h]
\vspace{5cm}
\caption{Evolution of the three coupling constants }
\end{figure}
\pagebreak

Recently, the more precise determination of the gauge coupling constants at
the scale $M_Z$ has allowed to show that, if only standard model particles
contribute to the Renormalization-Group Equations (RGE), the three running
coupling constants of $G\equiv SU(3)\otimes SU(2)\otimes U(1)$ meet at
three different points (Amaldi, de Boer, and Furstenau, 1991) [fig.1] and
only the meeting point of $\alpha_2(\mu)$
and $\alpha_S(\mu)$ corresponds to a value of the scale $\mu$ sufficiently
high to comply with the experimental lower limit on proton decay.

Indeed it has been observed (Amelino-Camelia, Buccella, and Rosa, 1990)
that $SO(10)$, for which the hypercharge is the
combination of two generators belonging to its Cartan,
\bee
Y=T_{3R}+\frac{B-L}{2},
\ene
is very promising to modify the SU(5) predictions in such a way to prevent
conflict with experiment. In fact, if there is an intermediate symmetry
group $G'$ containing $SU(2)_R$ and/or $SU(4)_{PS}$, to which
$T_{3R}$ and $B-L$ belong respectively, the non-Abelian evolution of either
component of Y implies a higher unification point.

An analysis of the possible symmetry breaking patterns with Higgses in
representations with dimension $\leq210$ and with only an
intermediate symmetry group $G'$ between $G$ and $SO(10)$ led to four
different possibilities for $G'$ (Buccella, 1988), which are reported in
table I together with the Higgs which must take the highest Vacuum
Expectation Value (VEV) ($\sim M_X$) in order to have spontaneous breaking
of $SO(10)$ to $G'$.$G'$ is then spontaneously broken to $G$ at the scale $M_R$
by an Higgs in the representation $126\oplus\overline{126}$.
In table I, $D$ is the left-right discrete symmetry, which interchanges
$SU(2)_L$ and $SU(2)_R$, first introduced by Kuzmin and Shaposnikov
(Kuzmin and Shaposhnikov, 1980).

\begin{table}
\centerline{TABLE I}
{\scriptsize
\begin{center}
\begin{tabular}{|c|c|c|} \hline
          &                             &   \\
  $G'$      & Higgs direction  & Representation   \\
          &          &        \\ \hline
          &                             &   \\
$\sq\times D$ &  $\omega_L=\frac{2\left(\omega_{11}+\ldots+\omega_{66}\right)
-3\left(\omega_{77}+\ldots \omega_{00}\right)}{\sqrt{60}}$   &  54 \\
          &                             &   \\ \hline
          &                             &   \\
$\sq   $  &  $ \gp_T= \gp_{7890}      $ & 210  \\
          &                             &   \\ \hline
          &                             &   \\
$\stt\times D$ & $\gp_L=\frac{\gp_{1234}+\gp_{1256}+\gp_{3456}}{\sqrt{3}}$ &
210\\
          &                              &  \\ \hline
          &                             &   \\
$\stt   $ & $\gp(\theta)=\cos\theta\gp_L+\sin\theta\gp_T$ & 210  \\
          &                              &  \\ \hline
\end{tabular}
\end{center}}
\caption{$\omega_{ab}$ is a second-rank traceless symmetric tensor;
$\gp_{abcd}$
is a fourth-rank antisymmetric tensor, and the indices 1...6 correspond to
$SO(6)\sim SU(4)_{PS}$, whereas 7...0 correspond to $SO(4)\sim
SU(2)_{L}\otimes SU(2)_{R}$.}
\end{table}

For the models in table I, using one-loop approximation for the RGE, one
obtains the following upper limit for the value of the scale of
$SO(10)$-breaking (Buccella, Miele, Rosa, Santorelli, and Tuzi, 1989):
\bee
M_X\leq M_Zexp{\pi\over2}{(\si-\aas)\over\alpha(M_Z)}.
\ene

An upper limit, that is more restrictive by a factor of about
$\frac{1}{3}$, is obtained using a two-loop approximation (Tuzi, 1989).

\begin{table}
\centerline{TABLE II}
{\scriptsize
\begin{center}
\begin{tabular}{|c|c|c|} \hline
          &                             &   \\
  $G'$      & $M_X/10^{15}~GeV$    & $M_R/10^{11}~GeV$  \\
          &          &        \\ \hline
          &                             &   \\
$\sq\times D$ &  $0.55\cdot1.6^{0\pm1}$   &  $340\cdot1.3^{0\pm1}$ \\
          &                             &   \\ \hline
          &                             &   \\
$\sq   $  &  $ 5.3\cdot1.9^{0\pm1} $ & $1.4\cdot2.1^{0\pm1}$     \\
          &                             &   \\ \hline
          &                             &   \\
$\stt\times D$ & $1.6\cdot2.8^{0\pm1}$  & $0.32\cdot1.8^{0\pm1}$     \\
          &                              &  \\ \hline
          &                             &   \\
$\stt   $ & $11\cdot2.1^{0\pm1}$    & $0.027\cdot3.3^{0\pm1}$    \\
          &                              &  \\ \hline
\end{tabular}
\end{center}}
\caption{Values of $M_X$ and $M_R$ for different $G'$ within the ESH.
Being unable to evaluate the errors at two-loops, here we quote
the errors obtained at one-loop; however, since the one-loop gives the main
contribution to RGE, we expect about the same order of magnitude for the
errors at one- and two-loops.}
\end{table}

The values of the scales $M_X$ and $M_R$ are reported in table II for the
different models within the Extended Survival Hypothesis (ESH) (Barbieri,
Morchio, Nanopoulos, and Strocchi, 1980),
namely allowing only the scalars needed for breaking the symmetry at the
lower scales to contribute to the RGE, and using the two-loop approximation
for the RGE.

As one can see in table II, both models without D symmetry give sufficiently
high values for the scale $M_X$, but the one with $G'\supset SU(4)_{PS}$
predicts $M_R \sim 10^{11}~GeV$, whereas the one with $G'\supset SU(3)_c\otimes
U(1)_{B-L}$ leads to a value of $M_R$ of about two orders of magnitude smaller.

At one-loop one finds an upper limit for $M_R$,
\bee
M_R\leq M_Z\exp{{\pi\over6\alpha(M_Z)}\left[\frac{3}{2}-3\si-\aas\right]},
\ene
where the equality holds only for the model with
$G'\supset SU(4)_{PS}\times D$, whose prediction for $M_X$ is too small.

The ESH may be too drastic since in the 210 and 126 representations of
$SO(10)$
there are multiplets with high quantum numbers, which may give important
contributions to the RGE beyond the ones implied by symmetry, the only ones
allowed within the ESH. This observation motivated Dixit and Sher to claim
that huge uncertainties are introduced in the $SO(10)$-predictions if the ESH
is removed (Dixit and Sher, 1989).
However, as explicitly shown in the following, the mass spectrum of the
scalars contributing to the RGE depends on
the coefficients of the non-trivial invariants which appear in the scalar
potential. These are constrained by demanding that the absolute minimum of
the potential is in the direction giving the desired
symmetry breaking pattern of the unified model considered.
This will allow us to deduce rather restrictive conditions on the
contributions of the scalars to RGE.

In this paper, we study the predictions for the scales
$M_X$ and $M_R$ obtained under the more general hypothesis  of allowing all the
scalars to contribute to the RGE between the
two highest scales, with their spectrum obeying the conditions necessary
for obtaining the symmetry breaking pattern characterizing each model.
By requiring for
$M_X$ a value sufficiently high to comply with the lower limit on $\t$ we shall
find upper limits on $M_R$, which correspond to lower limits on the masses of
the (almost) left-handed neutrinos.

\section*{II. Symmetry Breaking for Various Higgs Potentials}

\subsection*{A. $G' \equiv SU(4)_{PS}\otimes SU(2)_L\otimes SU(2)_R\times D$}

To construct a potential which gets its absolute minimum in the direction
corresponding to the desired symmetry breaking pattern, one can use non-trivial
positive definite invariants vanishing in that direction (Buccella and Ruegg,
1982; Kaymackcalan, Michel, Wali, Mc GLinn, and O'Raifeartaigh, 1986;
Burzlaff and O'Raifeartaigh, 1990). In order to have a renormalizable
potential the degree of these invariants should not exceed four.

In the case of the representation 54 (Shafi and Wetterich, 1979;
Buccella, Cocco, and Wetterich, 1984) (whose generic element we
denote by $\omega$) there is only one non-trivial quartic invariant,
\bee
||\left(\omega\omega\right)_{54}||=\sum_{i}(\omega\omega)_i(\omega\omega)_i
\ene
($(\omega\omega)_i\equiv\sum_{j,k} C^{54~54~54}_{~j~~k~~i}
\omega_j\omega_k,~~~(i,j,k=1\ldots54)$ and the $C$'s are $SO(10)$
Clebsh-Gordan),
which vanishes in the $SO(5)\otimes SO(5)$ direction, and is maximum in the
$SO(9)$-invariant one. To get the absolute minimum in the
$G_{422}\times D$-invariant direction $\omega_L$ one may consider the invariant
\bee
||\left(\omega\omega\right)_{54}+\alpha <\omega_L>\omega|| \label{eq:inv54}
\ene
($<\omega_L>$ is the VEV of $\omega_L$), vanishing in the $\omega_L$ direction
for $|\al|=\frac{1}{2\sqrt{42}}$ since
\bee
C^{54~~54~54}_{\omega_L~\omega_L~\omega}=-\frac{1}{2\sqrt{42}} \label{eq:cg54}
\dl_{\omega_L\omega}.
\ene

The invariant defined in eq. (\ref{eq:inv54}) gets its absolute minimum in
the $\omega_L$ direction (Abud, Buccella , Della Selva, Sciarrino, Fiore,
and Immirzi, 1986) for $|\al|$ in an open range around
$\frac{1}{2\sqrt{42}}$ with the lower and upper limits corresponding to the
degeneracy with the $SO(5)\otimes SO(5)$- or $SO(7)\otimes SO(3)$-invariant
directions respectively. This  implies a rather narrow range for the ratio of
the two masses m(1,3,3) and m(20,1,1) of the scalars of the 54
representation (Buccella and Rosa, 1992):

\bee  1.20<\frac{m(1,3,3)}{m(20,1,1)}< 1.35. \label{eq:rap54} \ene

Concerning the representation 126, a potential minimum in the SU(5)-invariant
direction $\p_0$, which is also the only $G$-invariant in the 126, can be
obtained (Buccella, Cocco, Sciarrino, and Tuzi, 1986) if the non-trivial part
of the potential is given by
($\p_+$ and $\p_-$ denote the 126 and $\overline{126}$ part respectively)
\beqa
V_\p(\p_+,\p_-) &=&
e_1||\cos{\theta_1}(\p_+\p_+)_{4125}+\sin{\theta_1}e^{i\eta_1}
(\p_-\p_-)_{4125}||+e_2||(\p_+\p_+)_{1050}|| \label{eq:vpsi} \\
&+& e_3||\cos{\theta_2}(\p_+\p_+)_{54}+\sin{\theta_2}e^{i\eta_2}
(\p_-\p_-)_{54}||, \nonumber
\enqa
where
\bee
(\p_{+\bl}\p_{+\bl})_{n_i}=C^{126~126~n}_{\bl~~~\bl~~~ i}
 <\p_+>^2,
\ene
with $\bl$ defining the vector $\p_+$ in the 126 representation.

The expression given in eq. (\ref{eq:vpsi}) is the most general combination of
the non-trivial invariants with the only limitation that the moduli of the
coefficients of $(\p_+\p_+\p_+\p_+)_1$ and of its Hermitean conjugate do
not exceed an appropriate combination of $|e_1|$ and $|e_3|$.

The positive-definite invariants appearing in eq. (\ref{eq:vpsi}) vanish in the
SU(5)-invariant direction $\p_0$ since
\bee
C^{126~126~r}_{\p_0~~\p_0~~i} \sim \dl_{r,2695}.
\ene
Clearly, if $e_i\geq0$ $V_\p$ has its absolute minimum in the SU(5)-invariant
direction.

In any case, these considerations are rather academic since the
direction of $\p_\pm$ is determined mainly by the non-trivial invariants
containing also $\omega$, which has a larger expectation value in the
scheme we are interested in. By considering the tensor products
\beqa
(54\otimes54)_s &=& 1\oplus54\oplus660\oplus770 \\
(126\otimes126)_s &=& 54\oplus1050\oplus2772\oplus4125 \nonumber \\
126\otimes\overline{126} &=&
1\oplus45\oplus210\oplus770\oplus5940\oplus8910 \nonumber
\enqa
(the subscript s means that one is considering the symmetric part of the
product of two identical representations) we observe that there is one
non-trivial invariant for the products
$\p_+\p_-(\omega \omega)_s$, $(\p_+\p_+)_s(\omega \omega)_s $ and
$(\p_+\p_+)_s\omega$, which we can write respectively as
\bee   ||(\p_+\omega )_{\overline{126}}||,~~(\p_+\p_+)_{54}\times(\omega
\omega)_{54} ~~{\rm and}~~(\p_+\p_+)_{54}\times \omega. \label{eq:invmis}
\ene

All the invariants in (\ref{eq:invmis}) vanish in the direction
$\omega_L,~\p_0$ since the 54 does not contains SU(5) singlets; hence
\bee
(\p_{+0}\p_{+0})_{54}=0
\ene
and
\bee
C^{126~ 54~\overline{126}}_{\p_0~~\omega_L~~i}=0
\ene
since $\p_0\omega_L$ is the $G$-singlet of a 24 of SU(5) and the only
$G$-singlet of the $\overline{126}$ is a singlet also of SU(5). Of the
three invariants defined in eq. (\ref{eq:invmis}) the first one
is positive-definite
while the other two change sign under the phase transformation
$\p_+\rw i\p_+$. Thus, to have the minimum in the desired direction it is
necessary to take a positive value for the coefficient of the first
invariant and to have sufficiently small coefficients for the other two,
which both tend to choose different directions for $\p_+$, once $\omega$
is fixed in the $\omega_L$ direction.

\subsection*{B. The 210 representation}

Because in all the other SO(10) models that we discuss the Higgs potential
depends on the reducible representation which is the sum of the $210,~126,~
\overline{126}$ and $10$ (whose generic elements we indicate with
$\gp,~~\p_+,~~\p_-,~~\rho$ respectively), we give a general analysis of the
orbit structure of this potential, which we indicate with
V($\gp,\p_+,\p_-,\rho)$ and can be written (Abud, Buccella, Rosa, and
Sciarrino, 1989) as
\beqa
V(\gp,\p_+,\p_-,\rho) &=& V_0(||\gp||,||\p||,||\rho||)+V_\gp(\gp)+
V_\p(\p_+,\p_-) \\
&+& V_{\gp,\p}(\gp,\p_+,\p_-)+V_{\gp,\p,\rho}(\gp,\p_+,\p_-,\rho), \nonumber
\enqa
where $V_0$ is a function of the fields' norms only, hence it is isotropic in
the space of representations and does not affect the direction of the
potential minimum.

The final little group is obtained from the
intersection of each of the little groups of the irreducible components of
the $210\oplus126\oplus\overline{126}\oplus10$. $V_\gp(\gp)$
breaks $SO(10)$ to $G'$, $V_\p(\p_+,\p_-)+V_{\gp,\p}(\gp,\p_+,\p_-)$
realizes the second symmetry breaking step to $G$ and finally
$V_{\gp,\p,\rho}(\gp,\p_+,\p_-,\rho)$ breaks down to
$SU(3)\otimes U(1)$.

Let $r,~ s$ and $t$ be the fields' norms in the vacuum state:
\bee
r=\sqrt{||\gp_0||},\ \ \ s=\sqrt{||\p_0||},\ \ \ t=\sqrt{||\rho_0||};
\ene
we can then write
\beqa
V_0(||\gp||,||\p||,||\rho||)&=& h_{\gp\gp}\left[||\gp||-r^2\right]^2+
h_{\p\p}\left[||\p||-s^2\right]^2+h_{\rho\rho}\left[||\rho||-t^2\right]^2 \\
&+&h_{\gp\p}\left[||\gp||-r^2\right]\left[||\p||-s^2\right]+h_{\gp\rho}
\left[||\gp||-r^2\right]\left[||\rho||-t^2\right] \nonumber \\
&+& h_{\p\rho}\left[||\p||-s^2\right]\left[||\rho||-t^2\right]. \nonumber
\enqa

Choosing $h$ so that we have a positive-definite quadratic form, $V_0$ is
minimum for
\bee
||\gp||=r^2,\ \ \ ||\p||=s^2,\ \ \ \ ||\rho||=t^2.
\ene

The remaining terms are ($V_\p$ is defined  in eq. (\ref{eq:vpsi}))
(Buccella, 1988):
\beqa
V_\gp(\gp) &=& A||(\gp\gp)_{45}||+B||(\gp\gp)_{54}||+C||(\gp\gp)_{210}||
\label{eq:vfi} \\
& &-r D\left[(\gp\gp)_{210}\times\gp\right]_1, \nonumber \\ & & \nonumber \\
V_{\gp,\p}(\gp,\p_+,\p_-) &=& f_1||\cos{\theta_3}(\gp\p_+)_{10}+
\sin{\theta_3}e^{i\eta_3}(\gp\p_-)_{10}|| \label{eq:vfipsi} \\
&+&f_2||\cos{\theta_4}(\gp\p_+)_{120}+\sin{\theta_4}e^{i\eta_4}(\gp\p_-)_{120}||
 \nonumber \\
&+&f_3||\cos{\theta_5}(\gp\p_+)_{320}+\sin{\theta_5}e^{i\eta_5}(\gp\p_-)_{320}||
 \nonumber \\
&+&f_4||(\gp\p_+)_{126}+k \p_+||+
f_5\left[(\p_+\p_-)_{45}\times(\gp\gp)_{45}\right]_1, \nonumber \\
 & & \nonumber \\
V_{\gp,\p,\rho}(\gp,\p_+,\p_-,\rho) &=& \left[P_{10}(\gp,\p_+,\p_-)\times\rho
\right]_1 \label{eq:vfipsiro} \\
&+&\left[\left(q_1(\gp\gp)_{54}+q_2(\p_+\p_+)_{54}+q^*_2(\p_-\p_-)_{54}
\right)\otimes(\rho\rho)_{54}\right]_1. \nonumber
\enqa
$(\gp\gp)_d$ is the representation of dimension $d$ in the product
$210\otimes210$, more precisely \linebreak
$(\gp\gp)_d\equiv \sum_i{C^{210~210~ d}_{\gp~~~\gp~~~i} d_i}$;
analogously for $(\p\p),~(\gp\p)$ and $(\rho\rho)$.$P_{10}$ is the most
general third order polynomial which transforms like a 10.

The expression of $V_{\gp,\p}$ given in eq. (\ref{eq:vfipsi}) contains
all non-trivial invariants that can be built with $\gp$ and $\p_{\pm}$,
with the only limitation that the coefficients of the invariants
$(\gp\p_\pm)_l\times(\gp\p_\pm)_l~~(l=10,120,320)$ have modulus sufficiently
small with respect to the coefficients of $||(\gp\p)_l||$.

The symmetry breaking pattern of the models that we analyze only differ in
the first step (they have different $G'$). We can therefore start by giving
a general discussion of the terms in the Higgs potential which are
responsible for the second and third symmetry breaking step. $V_\p$, as
defined in eq. (\ref{eq:vpsi}), gets certainly its minimum
in the SU(5)-invariant direction $\p_0$ if the $e_i$'s are all positive.
As far as $V_{\gp,\p}$ is concerned, if $\gp$ and $\p$ admit a common little
group $G$ then the terms $(\gp\p)_r$ with
$r\equiv10,~120,~320$ vanish because these representations do not contain
singlets under $G$ and therefore the first three terms in eq. (\ref{eq:vfipsi})
vanish in the $G$ direction.

The inhomogeneous term $||(\gp\p_+)_{126}+k\p_+||$ will vanish in  the same
direction if we take $\gp$ in a direction $\gp_0$ belonging to the
three-dimensional stratum invariant under $G$  and
\bee
k=-C^{210~126~126}_{\gp_0~~\p_0~~\p_0}\sqrt{||\gp_0||}.
\ene

In fact, taking $\gp$ and $\p$ singlets of $G$ the only non-vanishing
component of the product $\gp\p_+$ is in the direction $\p_0$. In conclusion,
taking $f_5$ small enough, $V_{\gp,\p}$ in eq. (\ref{eq:vfipsi}) will
be minimum
in the desired direction. To simplify our discussion we take $f_5=0$.

Let us finally consider $V_{\gp,\p,\rho}$.
Since, as we said, the 10 representation does not contain singlets under
$G$, $P_{10}$ vanishes at $\gp=\gp_0, \p=\p_0$. Since the 54 does not contain
SU(5) singlets, $(\p_+\p_+)_{54}$ and  $(\p_-\p_-)_{54}$ vanish in the
SU(5)-invariant direction. We have to choose the sign of $q_1$ in such a way
that $q_1 (\gp_0\gp_0)_{54}\times(\rho\rho)_{54}$
is smaller in the (1,2,2) $\rho$-direction (with respect to
$G_{422}$) rather than in the (6,1,1); with
our conventions for $SO(10)$ Clebsh-Gordan coefficients this happens if
\bee
q_1 C^{210~210~54}_{\gp_0~~\gp_0~~\omega_L} > 0, \label{eq:q1}
\ene
where $\gp_0$ is the direction of the 210 VEV and $\omega_L$ is the
$G_{422}$-invariant direction introduced
during the discussion of the breaking with the 54 and obeying eq.
(\ref{eq:cg54}).

In conclusion, allowing for $V_{\gp}(\gp)$ being minimum in a direction with
little group $G'$ containing $G$, we achieve our goal.

In the 210 representation there are three independent
$G$-singlets; therefore, the most general $G$-invariant direction lies on
the three-dimensional vector space (Abud, Buccella, Rosa, and Sciarrino,
1989)
\beqa
\gp_0(z_1,z_2,z_3) &=& z_1\frac{\gp_{1234}+\gp_{1256}+\gp_{3456}}{\sqrt{3}}
\label{eq:stratum3} \\
&+&z_2\frac{\gp_{1278}+\gp_{1290}+\gp_{3478}+\gp_{3490}+\gp_{5678}+\gp_{5690}}
{\sqrt{6}} \nonumber \\
&+&z_3\gp_{7890} \equiv z_1\gp_L + z_2\hat\gp + z_3\gp_T, \nonumber
\enqa
with $ z_1^2+z^2_2+z^2_3=1$.
For particular values of $z_i$, $\gp_0$ gets the following stability groups:
\[\bea{lcl}
{z_1}/{\sqrt{3}}={z_2}/{\sqrt{6}}=z_3 & \longrightarrow & SU(5)\otimes
U(1) \\    & & \\
z_1=z_2=0 & \longrightarrow & SU(4)_{PS}\otimes SU(2)_L\otimes SU(2)_R \\
& & \\
z_2=z_3=0 & \longrightarrow & SU(3)_C\otimes SU(2)_L\otimes SU(2)_R\otimes
U(1)_{B-L}\times D \\ & & \\
z_2=0,~|z_1|\neq0,1 & \longrightarrow & SU(3)_C\otimes SU(2)_L\otimes
SU(2)_R\otimes
U(1)_{B-L} \\ & & \\
\forall z_i & G' \supset & SU(3)_C\otimes SU(2)_L\otimes
U(1)_{T_{3R}}\otimes U(1)_{B-L}.
\ena\]

It is possible to show that in the 210 there are four non-trivial
independent invariants with degree$\leq 4$ which can be written as done in
eq. (\ref{eq:vfi}). For the phenomenological reasons mentioned in the
introduction,
we are interested in intermediate symmetries containing $SU(2)_R$ and (or)
$SU(4)_{PS}$, therefore we look for the minimum of $V_\gp(\gp)$ in the 2-
dimensional stratum characterized by the condition $z_2=0$. Writing the
expression of $V_\gp$ in this stratum ($\gp_0\equiv z_1\gp_L+z_3\gp_T$)
we get
\beqa
V_{2-dim.}(z_1,z_3)\equiv \frac{V_{\gp}(\gp_0(z_1,0,z_3))}{r^4} &=&
\frac{2}{35}A\left(z_1z_3\right)^2+
\frac{1}{210}B\left(2z_1^2-3z_3^2\right)^2 \label{eq:v2dim} \\
&+&\frac{2}{135}Cz_1^4-\frac{2}{3\sqrt{30}}Dz_1^3. \nonumber
\enqa

Because $z_1^2+z_3^2=1$ and using the new parameters
\bee
\al=\frac{4}{945}\left(-108A+225B+28C\right),\ \ \bl=\frac{2}{\sqrt{30}}D,\ \
\cl=\frac{4}{35}\left(2A-5B\right),\ \ \dl=\frac{3}{70}B \label{eq:newpar}
\ene
for later convenience, eq. (\ref{eq:v2dim}) becomes
\bee
V_{2-dim.}=\frac{\al}{8}z_1^4-\frac{\bl}{3}z_1^3+\frac{\cl}{4}z_1^2+\dl.
\ene

To have a relative minimum at $\tilde{z}_1\in]-1,1[$ we must
impose the conditions
\bee
\left(\frac{dV_{2-dim.}}{dz_1}\right)_{\tilde{z}_1}=0,\ \ \
\left(\frac{d^2V_{2-dim.}}{dz^2_1}\right)_{\tilde{z}_1}>0. \label{eq:minim}
\ene

The first condition is automatically satisfied at $z_1=0$
corresponding to the critical orbit with little group
$G_{422}$.

Moreover, we should have a positive-definite Hessian
\bee
\left.\frac{\partial^2 V_{\gp}(\gp)}{\partial\gp_i\partial\gp_j}
\right|_{\gp_0(\tilde{z}_1)}>0 \label{eq:hess}
\ene
in the subspace subtended by the scalars of the 210 surviving the Higgs
mechanism. The values of these masses, as a function of the parameters
defined in eq. (\ref{eq:newpar}) and of $z_1$ and $z_3$, are reported in the
Appendix, where one finds also the masses of the $126 \oplus
\overline{126}$ depending on the $f_i$'s defined in eq. (\ref{eq:vfipsi}).

Finally, we must be sure that $V_{2-dim.}(\tilde z_1)$ is an absolute
minimum in the 3-dimensional stratum $\gp_0(z_1,z_2,z_3)$, i.e.
\bee
V_{3-dim.}(z_1,z_2,z_3)\equiv \frac{V_\gp(\gp_0(z_1,z_2,z_3))}{r^4}
>V_{2-dim.}(\tilde{z}_1)~~~~~\forall (z_1,z_2,z_3)\neq (\tilde
z_1,0,\pm\sqrt{1-\tilde z_1^2}).
\label{eq:minass}
\ene

An indefinite metric for the Hessian would imply that $\gp_0(z_i)$ is not an
absolute minimum; hence the study of its positivity is a way to explore the
existence of $V(\gp)<V_{2-dim.}(\tilde{z}_1)$ in directions not belonging to
the $SU(3)\otimes SU(2)\otimes SU(2)\otimes U(1)\equiv G_{3221}$-invariant
stratum.

Using the following expression for $V_{3-dim.}(z_1,z_2,z_3)$,
\bee
V_{3-dim.}(z_1,z_2,z_3)=\frac{\al}{8}f_{\al}(z_i)-\frac{\bl}{3}f_{\bl}(z_i)+
\frac{\cl}{4}f_{\cl}(z_i)+\frac{\dl}{9}f_{\dl}(z_i),
\ene
\beqa
f_{\al}(z_i) &=& \left(z_1^2+z_2^2\right)^2+z^2_2\left(2z_1+\sqrt{z_3}\right)^2
+\frac{3}{4}z_2^4,  \\
f_{\bl}(z_i) &=& z_1^3+3z_1z^2_2+\sqrt{27/4}~z_2^2z_3, \\
f_{\cl}(z_i) &=& \left(z_1z_3+z_2^2/\sqrt{3}\right)^2+z_1^2z_2^2+\left(z_1^2+
z_2^2\right)^2+z_2^2\left(2z_1+\sqrt{3}z_3\right)^2 + \frac{3}{4}z^4_2, \\
f_{\dl}(z_i) &=& 30\left(z_1z_3+z_2^2/\sqrt{3}\right)^2+30z_1^2z_2^2+
\left(2z_1^2-z_2^2/2-3z_3^2\right)^2  \\
& & +5\left(z_1^2+z_2^2\right)^2 + 5z_2^2\left(2z_1+\sqrt{3}z_3\right)^2+
\frac{15}{4}z^4_2, \nonumber
\enqa
eq. (\ref{eq:minass}) becomes ($z_3=\sqrt{1-z_1^2-z_2^2}$)
\bee
f_\dl(z_1,z_2)\dl> f_\al(z_1,z_2)\al+ f_\bl(z_1,z_2)\bl+ f_\cl(z_1,z_2)\cl +
\sigma, \label{eq:mminass}
\ene
where $\sigma$ is a constant depending on $\tilde{z}_1$ and hence on the
residual symmetry of the model.
We distinguish the three regions represented in fig.2.

\begin{figure}
\vskip 5cm
\caption{{\(I_0=\left\{(z_1,z_2):f_\dl(z_1,z_2)=0\right\},~I_+=\left\{(z_1,z_2):
f_\dl(z_1,z_2)>0\right\},~I_-=\left\{(z_1,z_2):f_\dl(z_1,z_2)<0\right\}\) }}
\end{figure}

In order to satisfy the inequality (\ref{eq:mminass}) in each of the three
regions one needs that
\bee
f_\al\al+ f_\bl\bl+ f_\cl\cl + \sigma <0,~~~\dl_+<\dl<\dl_-,
\label{eq:reg}
\ene
with
\beqa
\dl_+ &\equiv& max\left\{\frac{\left[\al f_{\al}(z_i)+\bl f_{\bl}
(z_i)+\cl f_{\cl}(z_i) + \sigma \right]}{f_{\dl}(z_i)}
\right\}_{\left\{(z_1,z_2,z_3):f_{\dl}(z_i)>0\right\}} \\
\dl_- &\equiv& min\left\{\frac{\left[\al f_{\al}(z_i)+\bl f_{\bl}
(z_i)+\cl f_{\cl}(z_i) + \sigma \right]}{f_{\dl}(z_i)}
\right\}_{\left\{(z_1,z_2,z_3):f_{\dl}(z_i)<0\right\}}.
\enqa

By studying the last equation, combined with eq. (\ref{eq:hess}), we find that
it is sufficient to verify that the minimum obtained is lower than the value
corresponding to the direction invariant under the maximal little group
$SU(5)\otimes U(1)$, which lies in the $I_+$ region.

\subsubsection*{1. $G'\equiv SU(4)_{PS}\otimes SU(2)_L\otimes SU(2)_R$}

In Buccella, Cocco, Sciarrino, and Tuzi (1986) one has considered a
potential $V_\gp$ with positive
coefficients for the non-trivial invariants $||(\gp\gp)_{45}||,
{}~||(\gp\gp)_{210}||$ and $||(\gp\gp)_{1050}||$, which vanish in the
$G_{422}$-invariant direction. If the cubic term $(\gp\gp)_{210}\times\gp$
is absent or its coefficient is sufficiently small, the absolute minimum of
$V_\gp$ is in the $G_{422}$-invariant direction. In any case, a necessary
condition to have a minimum in the desired direction is
\bee  V(\gp_T)<V(\gp_L),  \label{eq:suftiz} \ene
which implies the following constraint on the mass spectrum:
\bee
\frac{2-\sqrt{3}}{2+\sqrt{3}} < \frac{m^2(15,3,1)}{m^2(15,1,3)} <
\frac{2+\sqrt{3}}{2-\sqrt{3}} . \label{eq:mastiz}
\ene

 From the identity
\bee
||(\gp\gp)_{54}||=-\frac{35}{14}||(\gp\gp)_{45}||-\frac{3}{7}||(\gp\gp)_{1050
}||
+\frac{15}{28}||(\gp\gp)_{210}||+\frac{1}{10}||\gp||^2
\ene
it is straightforward to translate the positivity of the coefficients of
$||(\gp\gp)_r||$ (r=45,210,1050) into the following inequalities for the
parameters defined in eq. (\ref{eq:newpar})
\bee
\cl>0,~~-\frac{27}{160} (\al+2\cl)<\dl<0, \label{eq:sufpar} \ene
while $\beta=0$ corresponds to the vanishing of the coefficient of the
cubic term.
If $\beta=0$ and the inequalities (\ref{eq:sufpar}) old, the absolute
minimum of $V_\gp$ is in the $G_{422}$-invariant direction $\gp_T$.
The necessary condition corresponding to the inequality (\ref{eq:suftiz}) reads
\bee
|\bl|< \frac{3}{8}(\al+2\cl).
\ene

 From eqs. (\ref{eq:minim}), (\ref{eq:hess}) and (\ref{eq:reg}) we find
the following necessary conditions to get the absolute minimum in the
$G_{422}$-invariant direction
\bee
\cl>0,~~\dl_T<\dl<0,~~\left\{\bea{ll}
-2\cl<\al\leq2\cl & 0\leq|\bl|<\frac{3}{8}(\al+2\cl) \\ &\\
\al>2\cl & 0\leq|\bl|<\frac{3}{2\sqrt{2}} \sqrt{\al\cl}  \ena \right. ,
\ene
where:
\bee
\dl_T\equiv \frac{1}{32\cdot15}\left(-81\al+24\sqrt{30}\bl-7\cdot27\cl\right).
\ene

\subsubsection*{2. $G'\equiv SU(3)_c\otimes SU(2)_L\otimes SU(2)_R\otimes
U(1)_{B-L}\times D$}

Two positive-definite invariants vanish in the $\gp_L$ direction
(Buccella and Rosa, 1987), i.e.
\bee
||\left(\gp\gp\right)_{45}||~~{\rm and}~~||\left(\gp\gp\right)_{210}
-C^{210~210~210}_{\gp_L~~\gp_L~~\gp_L}<\gp>\gp||. \label{eq:filu}
\ene

In fact, from the rule
\bee
C^{210~~210~~45}_{abcd~efgh~il}=\frac{1}{\sqrt{70}}\epsilon_{abcdefghil}
\ene
it follows that
\bee
\left(\gp_L\gp_L\right)_{45}=0
\ene
and $\gp_L$ enjoys the property of any critical direction
\bee
C^{210~210~210}_{\gp_L~~\gp_L~\gp_i}\sim \delta_{\gp_L\gp_i},
\ene
which implies a vanishing value for the second invariant defined by
eq. (\ref{eq:filu}).

Because other critical directions $\gp$, invariant under $G_{422}$ or
$SU(5)$ or $G(2)\otimes SU(2)$ (Baseq, Meljanac, and O'Raifeartaigh, 1989) give
rise to different values for $C^{210~210~210}_{\gp~~~\gp~~~\gp}$ the
conclusion is that the minimum of the potential is in the
$G_{3221}\times D$-invariant direction with positive coefficients for this
two invariants and smaller coefficients for
the others. With our parametrization the corresponding region is
\bee
\cl > 0,~~ \bl > 0,~~ -\frac{3}{40}\cl < \dl <0,~~ \{|\dl|,~|9\al + 18\cl
-12\bl
+40\dl|\}~{\rm ``small~ enough''}.
\ene

Again from eqs. (\ref{eq:minim}), (\ref{eq:hess}) and (\ref{eq:reg}) we find
that the following are the necessary conditions to get the absolute minimum in
the direction $\gp_L$:
\bee
\left.\bea{l}
\al > 0,~~\left\{\bea{ll}
-\frac{\al}{4} < \cl < \frac{\al}{2} & \frac{\al+\cl}{2}
\leq \bl < \frac{2\al+5\cl}{2}  \\ & \\
\cl \geq \frac{\al}{2} & \frac{3}{8}(\al+2\cl) < \bl <
\frac{2\al+5\cl}{2} \\  & \\
\ena \right.  \\ \\
\al \leq 0 \left.\bea{ll}
\cl > -\frac{\al}{2} & \frac{3}{8}(\al+2\cl) < \bl < \frac{5}{4}(\al+2\cl)
\ena \right.
\ena\right\} \dl_1 < \dl < \dl_2, \label{eq:condlu}
\ene
where
\beqa
\dl_1 &\equiv& max\left\{\frac{-5\al+4\bl-10\cl}{20},
\frac{9(-\al+\bl-\frac{5}{2}\cl)}{80},
\frac{-51\al+8\left(3\sqrt{30}-10\right)\bl
-129\cl}{32\cdot 15}\right\} \\
\dl_2 &\equiv& min\left\{0,\frac{1}{20}\left(-3\al+6\bl-5\cl\right)\right\}.
\enqa

\subsubsection*{3. $G'\equiv SU(3)_c\otimes SU(2)_L\otimes SU(2)_R\otimes
U(1)_{B-L}$}

The $G_{3221}$-invariant stratum can be described, as long as
$\sin 2\theta\neq 0$, by
\bee
\gp(\theta)=\cos\theta\gp_L+\sin\theta\gp_T. \label{eq:abrs}
\ene

Abud et al. (Abud, Buccella, Rosa, and Sciarrino, 1989)
considered the symmetry breaking direction
$\gp(\theta^*)$ with $\theta^* \equiv arcos\sqrt{\frac{3}{5}}$.
By exploiting the fact that $||[\gp(\theta^*)\gp(\theta*)]_{54}||=0$,
they found sufficient conditions to get $G'=G_{3221}$. However,
as already pointed out in Abud, Buccella, Rosa, and Sciarrino (1989),
the direction $\gp(\theta^*)$
does not obey the inequality (\ref{eq:q1}) necessary to construct a
$V_{\gp=\gp_0,\p,\rho}$ breaking $SU(2)_L$.

A search for different directions in the $\gp(\theta)$-stratum has
been performed in the theses of two of us (Amelino-Camelia, 1989;
Pisanti, 1992).
In particular, the challenging program of finding all the
conditions on the potential parameters necessary to get the minimum in
the $\gp(\theta)$-stratum has been carried out in Pisanti (1992).

By imposing that $V_\gp$, restricted on the 3-dimensional stratum defined
by eq. (\ref{eq:stratum3}), has its absolute minimum at $z_2=0$ and
$z_1 z_3\neq0$, and by
requiring the positivity of the mass spectrum, which gives some confidence
in the fact that the absolute minimum of $V_\gp$ does not lie in a direction
different from the ones defined by eq. (\ref{eq:abrs}), we get four cases,
all obeying the conditions
\bee
\al>0,~~\dl_{SU(5)}<\dl,
\ene
($\dl_{SU(5)}$ is the value of $\dl$ for which there is a
degeneracy with the $SU(5)\otimes U(1)$-invariant direction),where
\beqa
\dl_{SU(5)} &\equiv& \frac{1}{32\cdot15}\left(-81\al+24\sqrt{30}\bl-7\cdot27\cl
+\frac{10}{3}\sigma\right) \\
\sigma &\equiv& -\frac{3}{\al^3}\left[8\bl^4+3\al^2\cl^2-12\al\bl^2\cl
+8\bl\left(\bl^2-\al\cl\right)^{\frac{3}{2}}\right]
\enqa
and with the further constraints on the parameters described in table III.

\centerline{TABLE III}
\medskip

$ \bea{|c|ccc|} \hline
 & & & \\
A) &0<\cl<\frac{\al}{2}, & \frac{3}{\sqrt{8}}\sqrt{\al\cl}<\bl<\frac{\al+\cl}
{2}, & \dl< min\left\{\dl_{322},\dl_{813}\right\} \\
& & & \\ \hline
& & & \\
B) &-\frac{3}{5}\al<\cl\leq-\frac{1}{2}\al, & 0<\bl<\bl_1(\cl) &
\dl_{313}<\dl<min\left\{\dl_{322},\dl_{622},
\dl_{813}\right\} \\
 & & & \\
 & -\frac{1}{2}\al<\cl\leq\cl_1, &  \frac{\al+2\cl}{\sqrt{8}}\sqrt{\frac{-\cl}
{\al-2\cl}}<\bl<\bl_1(\cl) &      ''    \\
& & & \\
 & \cl_1<\cl\leq\cl_2, & \frac{\al+2\cl}{\sqrt{8}}\sqrt{\frac{-\cl}
{\al-2\cl}}<\bl<\bl_2(\cl),  &    ''  \\
& & & \\
 & ''&\bl_4(\cl)<\bl<\frac{\al+\cl}{2}, & \dl<\dl_{313} \\
& & & \\
 & \cl_2<\cl\leq\cl_3, & \frac{\al+2\cl}{\sqrt{8}}\sqrt{\frac{-\cl}{\al-2\cl}}<
\bl\leq
\frac{12\al+13\cl}{4\sqrt{39}}, & \dl<min\{\dl_{322},\dl_{622},
\dl_{813}\} \\
& & & \\
 & '' & \left\{\bea{c}
\frac{12\al+13\cl}{4\sqrt{39}}<\bl<\bl_3(\cl)  \\   \\
\bl_4(\cl)<\bl<\frac{\al+\cl}{2}, \ena \right.  &
\dl<min\{\dl_{322},\dl_{313}\} \\
& & & \\
 & \cl_3<\cl<0 & \frac{\al+2\cl}{\sqrt{8}}\sqrt{\frac{-\cl}{\al-2\cl}}<\bl\leq
\frac{12\al+13\cl}{4\sqrt{39}}, & \dl<min\{\dl_{322},\dl_{622},
\dl_{813}\}   \\
& & & \\
 &'' & \frac{12\al+13\cl}{4\sqrt{39}}<\bl<\frac{\al+\cl}{2}, &
\dl<min\{\dl_{322},\dl_{313}\} \\
 & & & \\ \hline
 & & & \\
C) & -\frac{3}{5}\al<\cl<-\frac{1}{2}\al & \bl=0 & \dl<\dl_{622}\\
& & &\\  \hline
 & & & \\
D) & \cl=0 & \left\{\bea{c}
 0<\bl\leq\bl_-  \\ \bl_+\leq\bl<\frac{1}{2}\al \ena,\right. &
\dl<0\\
 & &\bl_-<\bl<\bl_+ & \dl<\dl_{813} \\
& & & \\
\hline
\ena $
\bigskip

In table III, $\dl_i$ is the value of $\dl$ for which $m_i$ vanishes
(e.g. $m(3,2,2,-2/3)=0$ for $\dl=\dl_{322}$);
moreover:
\begin{itemize}
\item ${\displaystyle
\cl_1=\frac{-91+12\sqrt{30}}{169-12\sqrt{30}}\al\simeq-0.245\al}$
\item $\cl_2=-3/19\al$
\item $\bl_3(\cl),~~\bl_4(\cl)$ are defined by the
equality $\dl_{313}(\bl_{3,4},\cl)=\dl_{SU(5)}(\bl_{3,4},\cl)$
\item $\bl_1(\cl)$ is such that $\dl_{313}(\bl_1,\cl)=\dl_{813}(\bl_1,\cl)$
\item $\bl_2(\cl)$ is such that
$\dl_{313}(\bl_2,\cl)=min\left\{\dl_{322}(\bl_2,\cl),
\dl_{813}(\bl_2,\cl)\right\}$
\item $\cl_3$ is defined by $\bl_3(\cl)=
\bl_4(\cl)$
\item $\dl_{322}=-\cl/10$
\item ${\displaystyle
\dl_{622}=\frac{3\left[-2\bl^2\cl+2\al\bl^2+\al\cl^2+2\bl(\al-\cl)
\sqrt{\bl^2-\al\cl}\right]}
{10\left[3\al^2-2\bl^2+\cl-2\bl)\sqrt{\bl^2-\al\cl}\right]}}$
\item ${\displaystyle
\dl_{313}=9\frac{\frac{3}{4}\left(\al+2\cl\right)-\bl z_1-\sqrt{3}\bl z_3-
\frac{9\al+54\cl}{36}z_1^2+\frac{9\al+21\cl}{\sqrt{108}}z_1z_3+\bl z_1^3-
\frac{\al}{2}z_1^4}{20z_1\left(z_1-2\sqrt{3} z_3\right)}}$
\item ${\displaystyle
\dl_{813}=9\frac{\frac{3}{4}\left(\al+2\cl\right)+\bl z_1-\sqrt{3}\bl z_3-
\frac{27\al+90\cl}{36}z_1^2-\frac{9\al+21\cl}{\sqrt{108}}z_1z_3+\bl z_1^3-
\frac{\al}{2}z_1^4}{40z_1\left(z_1+\sqrt{3} z_3\right)}}$
\item $\bl_{\pm}=\frac{\sqrt{27\pm12\sqrt{2}}}{14}\al\simeq\left\{\bea{c}
0.4756\al \\
\\ 0.2262\al~~. \ena\right.$
\end{itemize}

\section*{III. Renormalization-Group Equations}

In the following we study the implications of the RGE for each of the
SO(10) models that we have considered.
In the RGE-analysis presented in Mohapatra and Parida (1993)
and Deshpande, Keith, and Palash Pal (1992) the simplifying
ESH has been assumed. We shall remove this assumption,
and consider all possible scalar contributions to the RGE above $M_R$ that are
compatible with the necessary conditions which we obtained in the preceding
section by demanding that the absolute minimum of
the potential is in the direction giving the desired symmetry breaking
pattern.

The evolution of the coupling constants $\alpha_i(\mu)$ is governed by the
Gell-Mann Low equation
\bee
\mu\frac{d}{d\mu}\alpha_i(\mu) = \beta_i(\alpha_i(\mu)),
\ene
with
\bee
\beta_i(\alpha_i(\mu)) = a_i\alpha^2_i(\mu)+\sum_j b_{ij}\alpha_j(\mu)
\alpha^2_i(\mu)+\ldots
\ene
(no sum on the repeated indices).

The general expression for $a_i$ and $b_{ij}$ for a gauge group $G=G_1\otimes
G_2$ is (Jones, 1982)
\beqa
a_i &=& \frac{1}{2\pi}\left[\frac{2}{3}\sum_F T(F_i)F_j + \frac{1}{6}\sum_S
T(S_i)S_j - \frac{11}{3} C_2(G_i) \right]  ~~(i=1,2) \\  & & \nonumber  \\
b_{ij} &=& \frac{1}{8\pi^2}\left[2\sum_F C_2(F_j)F_jT(F_i) + 2\sum_S
C_2(S_j)S_jT(S_i) \right]  ~~(i,j=1,2;~i\neq j) \\
& & \nonumber     \\
b_{ii} &=& \frac{1}{8\pi^2}\left[\sum_F \left(\frac{10}{3}C_2(G_i) + 2C_2(F_i)
\right)F_jT(F_i) \right] \\
 &+& \frac{1}{8\pi^2}\left[\sum_S \left(\frac{1}{3}C_2(G_i) + 2C_2(S_i)
\right)S_jT(S_i) - \frac{34}{3} \left(C_2(G_i)\right)^2 \right] ~~~(i=1,2).
\nonumber
\enqa

\begin{itemize}
\item $F_i$ and $S_i,~(i=1,2)$ are the dimension of the multiplet
that classifies
fermions and scalars respectively under $G_1\otimes G_2$.
\item $C_2(F)$ is the second-order Casimir of the representation F.
\item $G_i$ is the adjoint representation.
\item $T(F)\equiv\frac{C_2(F)F}{dim(adjoint)}.$
\item If $G= U(1)$ with generator $\sigma$ then
$C_2(G)=0$, $C_2(F)=\frac{1}{2} Tr(\sigma(F)^2)=T(F)$.
\end{itemize}

At the scales $M_I$ of the spontaneous symmetry breaking we impose for the
coupling constants, which for symmetry reasons are equal above $M_I$ whereas
they evolve independently below $M_I$, the matching conditions
(Weinberg, 1980; Mohapatra and Parida, 1993)
\bee
\frac{1}{\al_i(M_I)}=\frac{1}{\al_I}-\frac{C_2(G_i)}{12\pi}.
\ene

\subsection*{A. $G' \equiv SU(4)\otimes SU(2)\otimes SU(2)\times D$}

In writing the RGE we define the two parameters $A$
and $B$ as
\beqa
A~ &\equiv& ~{6\pi\over 11\alpha(M_Z)}\left(\si-\aas\right) \\
B~ &\equiv& ~{\pi\over 11\alpha(M_Z)}\left({3\over 2}-3\si-\aas\right),
\enqa
where (Anselmo, Cifarelli, and Zichichi, 1992)
\begin{itemize}
\item $\al(M_Z)=\frac{1}{127.9\pm 0.2}$
\item $\al_S(M_Z)=0.118\pm 0.008$
\item $\si=0.2334\pm 0.0008$,
\end{itemize}
and, for example, we get at one-loop
\beqa
A-{12\over11}B &=& {25\over11}\ln{M_X\over M_R}+{1\over11}\Biggr[
-\ln{M_X\over m(15,2,2)_{126}}-\ln{M_X\over m(6,1,1)_{126}} \label{eq:ablor} \\
&-& 4\ln{M_X\over m(20,1,1)_{54}}+3\ln{M_X\over m(1,3,3)_{54}}-
\ln{M_X\over m(6,1,1)_{10}}\Biggr] \nonumber \\
M_R &=& M_Z\exp{B}=M_{SU(5)} \label{eq:rlor}
\enqa
(in these equations we insert all the scalars but, according to the
Appelquist-Carazzone theorem (Appelquist and Carazzone, 1975), they
contribute only if their
mass is below $M_X$), where the coefficient of the first term on the
r.h.s. of eq. (\ref{eq:ablor}) is given by the
sum of the contributions of the gauge bosons (2), the Higgs doublets of the 10
$(2{1\over22})$\footnote{We take two 10 representations for the Higgses
responsible of the spontaneous breaking of the electroweak theory to avoid the
prediction $m_t=m_b$. The discussion on $V_{\gp,\p,\rho}$ of the
previous section
may be easily generalized to the case of two 10's.}
and the triplets of the $126\oplus\overline{126}$\ under
$SU(2)_R$\ (1) or $SU(2)_L ({-9\over11})$.
By neglecting the term in square brackets, absent within ESH, we get at
one- and two-loops respectively\footnote{The values at two-loops are the
result of a numerical analysis.}:
\[\bea{lll}
M^{(1)}_R=4.0\cdot10^{13}\cdot1.3^{0\pm1} & M^{(1)}_X =1.0\cdot10^{15}
\cdot1.6^{0\pm1} & {\t^{(1)}} =1.1\cdot10^{31}\cdot1.6^{0\pm4} \\
M^{(2)}_R=3.4\cdot10^{13} & M^{(2)}_X=5.5\cdot10^{14} &
\t^{(2)}=8.6\cdot10^{29},
\ena\]
in disagreement with the lower limit $\t\geq9\cdot10^{32}~years$.

Since $M_R$ is fixed by eq. (\ref{eq:rlor}), in order to get the highest
possible value for $M_X$ one would like to have the highest possible value
($\sim M_X$) for the (1,3,3) multiplet appearing on the r.h.s. of eq.
(\ref{eq:ablor}) and the lowest possible value ($\sim M_R$) for the remaining
masses. However, if the absolute minimum of
the scalar potential built with the 54 is in the $G_{422}\times D$-invariant
direction, eq. (\ref{eq:rap54}) implies $m(1,3,3)< 1.35~m(20,1,1)$
(Buccella and Rosa, 1992), and consequently the contribution of the
second term in the r.h.s. of eq. (\ref{eq:ablor}) would be
$> -\frac{4}{11}\ln{\frac{M_X}{M_R}}-\frac{3}{11}\ln{1.35}$. One would then
get at one-loop
\bee
M_X<M_Z ~ e^{{1\over 21}(11A+9B+3\ln{1.35})}
= (2.0\cdot10^{15}\cdot1.8^{0\pm1})~GeV.
\ene

With all the multiplets at the scale $M_R$ but the (1,3,3) one, for which
we have $m(1,3,3)\leq 1.35~M_R$, one would get:
\[\bea{lll}
M^{(1)}_R=4.0\cdot10^{13}\cdot1.3^{0\pm1} & M^{(1)}_X = 2.0\cdot10^{15}
\cdot1.8^{0\pm1} & \t^{(1)} = 1.5\cdot10^{32}\cdot1.8^{0\pm4} \\
M^{(2)}_R = 3.6\cdot10^{13} &  M^{(2)}_X = 1.1\cdot10^{15} &
\t^{(2)} = 1.2\cdot10^{31}.
\ena\]

$M^{(2)}_X$\ is too small (about two standard deviations) to comply with the
lower limit in eq. (\ref{eq:mxinf}).

\subsection*{B. $G'\equiv SU(4)_c\otimes SU(2)_L\otimes SU(2)_R$}

In this case, one finds
\beqa
B &=& \ln{M_R\over M_Z}+\frac{5}{44}\Biggr[
3\ln{\frac{m(15,3,1)}{m(15,1,3)}}+4\ln{\frac{m(10,3,1)}{m(\overline{10},1,3)}}
\Biggr] \label{eq:btiz} \\
A+2B &=&
\frac{\pi}{11}\left(\frac{3}{\alpha(M_Z)}-\frac{8}{\alpha_S(M_Z)}\right)=
\label{eq:abtiz} \\
&=& 3\ln{M_R\over M_Z}+2\ln{M_X\over M_R}
+{1\over11}\left[\ln{\frac{m(6,1,1)_{10}}{m(1,2,2)_{10}}}
+\frac{3}{2}\ln{\frac{M_X^2}{m(15,3,1)_{210}m(15,1,3)_{210}}}
\right. \nonumber \\
&+& \ln{\frac{M_X^2}{m(10,3,1)_{126}m(\overline{10},1,3)_{126}}}
-2\ln{\frac{M_X}{m(15,1,1)_{210}}}-2\ln{\frac{M_X}{m(10,2,2)_{210}}}
\nonumber \\
&-& \left. \ln{\frac{M_X}{m(15,2,2)_{126}}}-\ln{\frac{M_X}{m(6,1,1)_{126}}}
\right]. \nonumber
\enqa

The ESH would imply $m(1,2,2)_{10}\sim M_Z$, $m(\overline{10},1,3)\sim M_R$ and
all the other scalars with mass $\sim M_X$. One would obtain at one- and
two-loops respectively:
\[\bea{lll}
M_R^{(1)}=5.2\cdot10^{11}\cdot2.1^{0\pm1} & M^{(1)}_X=7.1\cdot10^{15}
\cdot1.9^{0\pm1} & \t^{(1)}=2.5\cdot10^{34}\cdot1.9^{0\pm4} \\
M^{(2)}_R=1.4\cdot10^{11} & M^{(2)}_X=5.3\cdot10^{15} &
\t^{(2)}=7.6\cdot10^{33}.
\ena\]

The value found for $M_R$ is lower than $M_{SU(5)}$ and it is
phenomenologically intriguing
(Mohapatra and Parida, 1993; Babu and Shafi, 1993).
To establish how much the
value found depends on ESH, we look for the highest value for $M_R$
consistent with eq. (\ref{eq:mxinf}).

 From eqs. (\ref{eq:btiz}) and (\ref{eq:mastiz}), by taking $m(10,3,1)\geq
M_R$,
one gets
\bee
M_R<M_Z~ e^B~ e^{\frac{15}{88}\ln{\frac{2+\sqrt{3}}{2-\sqrt{3}}}}
= (6.2\cdot10^{13}\cdot1.4^{0\pm1})~GeV.
\ene
$M_R$ takes the value $6.2\cdot10^{13}~GeV$ if
\beqa
m(10,3,1)_{210} &=& m(\overline{10},1,3)_{210} \label{eq:mtiz1} \\
m^2(15,3,1)_{210} &=& \frac{2-\sqrt{3}}{2+\sqrt{3}}m^2(15,1,3)_{210},
\label{eq:mtiz2}
\enqa
which corresponds to $\bl=-\frac{3}{8}(\al+2\cl)$, just on the boundary of
the allowed values for $\bl$ that are compatible with the absolute minimum
being in the $G_{422}$-invariant direction.
For that value of $\bl$, $\al$ should be $\leq 2\cl$ and one has the
inequality
\bee
\frac{m^2(15,3,1)_{210}m^2(15,1,3)_{210}}{m^4(15,1,1)_{210}}=
\frac{9(\al+2\cl)^2}{16\cl^2}
\leq 9. \label{eq:distiz}
\ene

 From eqs. (\ref{eq:abtiz}), (\ref{eq:mtiz1}), (\ref{eq:mtiz2}), and
(\ref{eq:distiz}) one gets:
\bee
M_R \leq 3^{\frac{1}{14}} M_Z \left(\frac{M_Z}{M_X}\right)^{\frac{10}{7}}
e^{\frac{\pi}{14}\left(\frac{3}{\alpha(M_Z)}-\frac{8}{\alpha_S(M_Z)}\right)}
\leq 3\cdot10^{13}~GeV,
\ene
the last inequality coming from eq. (\ref{eq:mxinf}).

The highest value for $M_R$ consistent with eq. (\ref{eq:mxinf})
is found by taking $(1,2,2)_{10}$ at the scale
$M_Z$, the multiplets $(10,2,2)_{210},~(6,1,1)_{10}$ and all the states of
the 126 at the scale $M_R$ and all the other states of the 210 at the scale
$M_X$; in such conditions the numerical analysis gives:
\[\bea{lll}
M_R^{(1)}=1.6\cdot10^{13}\cdot2.5^{0\pm1} & M^{(1)}_X=4.7\cdot10^{15}
\cdot1.8^{0\pm1} & \t^{(1)}=4.8\cdot10^{33}\cdot1.8^{0\pm4} \\
M^{(2)}_R=1.1\cdot10^{13} & M^{(2)}_X=3.1\cdot10^{15} &
\t^{(2)}=9.2\cdot10^{32}.
\ena\]

\subsection*{C. $G'\equiv SU(3)_c\otimes SU(2)_L\otimes SU(2)_R\otimes
U(1)_{B-L}\times D$}

In this case, by keeping into account the $SU(2)_L\leftrightarrow SU(2)_R$
symmetry above $M_R$, we get

{\footnotesize
\beqa
2B-A &=& \ln{M_R\over M_Z}+{1\over11}\left[-\ln{M_X\over m(1,2,2,0)_{10}}
+ \ln{\frac{M_X}{m(1,3,1,-2)_{126}}}
+ 2 \ln{\frac{m(8,3,1,0)_{210}}{m(6,2,2,2/3)_{210}}} \right.\label{eq:comblu}
\\
&+& \ln{\frac{M_X}{m(1,2,2,2)_{210}}}
+ 2\ln{\frac{m(3,3,1,-2/3)_{126}}{m(3,2,2,4/3)_{126}}}
+ \ln{\frac{m(3,3,1,-2/3)_{126}}{m(3,1,1,-2/3)_{126}}} \nonumber \\
&+& \left. \ln{m(3,2,2,-2/3)_{210}\over m(8,1,1,0)_{210}}
+ \ln{m(1,2,2,0)_{126}\over m(3,1,1,-2/3)_{126}} \right] \nonumber \\
B &=& \frac{1}{2} \ln{\frac{M_X M_R}{M_Z^2}}
+{1\over11}\left[ {3\over2}\ln{\frac{M_X}{m(1,3,1,-2)_{126}}}
+ \ln{\frac{m(6,2,2,2/3)_{210}}{m(1,2,2,2)_{210}}}
\right. \label{eq:blu} \\
&+& {3\over2}\ln{\frac{m(8,3,1,0)_{210}}{m(3,3,1,4/3)_{210}}}
-{1\over4}\ln{M_X\over m(8,1,1,0)_{210}}
+2\ln{\frac{m(8,2,2,0)_{126}}{m(3,2,2,4/3)_{126}}} \nonumber \\
&-& \left. {3\over2}\ln{\frac{M_X}{m(6,3,1,2/3)_{126}}} \right]. \nonumber
\enqa}

The ESH would imply $m(1,2,2,0)_{10}\sim M_Z$,
$m(1,3,1,-2)_{126}\sim M_R$ and all
the other multiplets at the scale $M_X$. One should get at one- and two-loops
\[\bea{lll}
M_R^{(1)}=1.2\cdot10^{10}\cdot1.8^{0\pm1}& M^{(1)}_X=4.1\cdot10^{15}
\cdot2.8^{0\pm1} & \t^{(1)} =2.8\cdot10^{33}\cdot2.8^{0\pm4} \\
M^{(2)}_R=3.2\cdot10^{10} & M^{(2)}_X=1.6\cdot10^{15} &
\t^{(2)}=7.0\cdot10^{31},
\ena\]
a too small value for $\t^{(2)}$.

In considering the contribution of the other scalars we take into account
the constraints on the mass spectrum, which follow from the requirement
that the absolute minimum falls in the desired direction. The parameters
defined in (\ref{eq:newpar}) should obey eq. (\ref{eq:condlu}), which
implies the following inequalities for the masses (the case discussed here
is obtained with $z_1=1~z_3=0$):
\bee
\frac{m(8,3,1,0)_{210}}{m(3,3,1,4/3)_{210}}>\sqrt{\frac{37}{14}}
,~~\frac{m(6,2,2,2/3)_{210}}{m(1,2,2,2)_{210}}
>\frac{1}{\sqrt{7}},~~
1<\frac{m(8,3,1,0)_{210}}{m(6,2,2,2/3)_{210}}<\frac{2}{\sqrt{3}}.
\label{eq:masluig}
\ene

 From the positivity of the $f_i$'s defined in eq.(\ref{eq:vfipsi}) and from
the expressions for the masses in the Appendix one gets also
\bee
1\leq \frac{m(3,3,1,-2/3)_{126}}{m(3,2,2,4/3)_{126}}\leq 2 \label{eq:buc}.
\ene

 From eqs. (\ref{eq:comblu}), (\ref{eq:blu}), (\ref{eq:masluig}), and
(\ref{eq:buc}) one finds the inequalities
\beqa
M_R &\leq& \left(\frac{2\cdot 7^{\frac{5}{3}}}{37}\right)^{\frac{3}{31}}
\frac{M_{SU(5)}^{\frac{44}{31}}}{M_X^{\frac{13}{31}}} \label{eq:mrfir} \\
M_R &\leq& M_Z^{\frac{5}{6}}M_X^{\frac{1}{6}}
e^{{\pi\over \alpha(M_Z)}\left({1\over 4}-\si+\frac{1}{3}\aas\right)}.
\label{eq:mrsec}
\enqa

The highest value for $M_R$ is found by taking $m(8,1,1,0)_{210}$,
$m(6,3,1,2/3)_{126}$ and \linebreak
$m(8,2,2,0)_{126}$ at the scale $M_R$ and
$m(3,2,2,4/3)_{126}$ at the scale $M_X$ for eq. (\ref{eq:mrfir}), and
$m(3,2,2,-2/3)_{210}$, $m(3,3,1,-2/3)_{126}$, and $m(3,2,2,4/3)_{126}$
at the scale $M_R$, $m(1,2,2,2)_{210}$, $m(8,1,1,0)_{210}$,
$m(3,1,1,-2/3)_{126}$, and $m(1,2,2,0)_{126}$ at the scale $M_X$,
and $m(8,3,1,0)_{210}=m(6,2,2,2/3)_{210}$ for eq. (\ref{eq:mrsec}).
The two requirements may not be satisfied at the same time, since they
imply a different scale for $m(8,1,1,0)_{210}$ and $m(3,2,2,4/3)_{126}$.
Except for the factor $\left(\frac{2\cdot 7^{\frac{5}{3}}}{37}\right)
^{\frac{3}{31}}
\sim 1.03$, eq. (\ref{eq:mrfir}) is the same found in Buccella and Rosa (1992)
with more restrictive conditions.

By eliminating $M_X$ in eqs. (\ref{eq:mrfir}) and (\ref{eq:mrsec}) one
finds the inequality
\bee M_R<3.1\cdot10^{11}\cdot 3.1^{0\pm1}~GeV. \label{eq:buc1} \ene

Certainly, it would be possible to get a lower bound for $M_R$ since the
one just written has been obtained by multiplying inequalities which
cannot be both equalities. So we are not surprised when,
by looking for the highest value for $M_R$ consistent with eq.
(\ref{eq:mxinf}) and with the constraints on the spectrum
following from eq. (\ref{eq:condlu}), the numerical analysis gives:
\[\bea{lll}
M_R^{(1)}=2.4\cdot10^{10}\cdot3.1^{0\pm1}& M^{(1)}_X=8.1\cdot10^{15}
\cdot1.9^{0\pm1} & \t^{(1)} =4.1\cdot10^{34}\cdot1.9^{0\pm4} \\
M^{(2)}_R=2.9\cdot10^{10} & M^{(2)}_X=3.1\cdot10^{15} &
\t^{(2)}=9.3\cdot10^{32},
\ena\]
with $M_R^{(1)}$ lower than the r.h.s. of eq. (\ref{eq:buc1}).

\subsection*{D. $G'\equiv SU(3)_c\otimes SU(2)_L\otimes SU(2)_R\otimes
U(1)_{B-L}$}

In this case, the absence of D symmetry brings to more complicate expressions
for A and B, namely
{\footnotesize
\beqa
A &=& \ln{M_X\over M_Z}+{1\over11}\left[\ln{M_X\over m(1,2,2,0)_{10}}
+ \ln{\frac{m(8,3,1,0)_{210}}{m(3,3,1,4/3)_{210}}}
+ \ln{\frac{m(6,2,2,2/3)_{210}}{m(1,2,2,2)_{210}}} \right. \\
&+& 2\ln{\frac{m(6,2,2,2/3)_{210}}{m(3,3,1,4/3)_{210}}}
+ \ln{\frac{m(6,2,2,2/3)_{210}}{m(3,2,2,-2/3)_{210}}}
- {3\over2}\ln{M_X\over m(8,1,1,0)_{210}} \nonumber \\
&+& 2\ln{\frac{M_X}{m(1,3,1,-2)_{126}}}
- \ln{M_X\over m(3,1,1,-2/3)_{126}}
+ 2\ln{\frac{m(8,2,2,0)_{126}}{m(3,2,2,4/3)_{126}}} \nonumber \\
&+& \left. 3\ln{\frac{m(6,3,1,2/3)_{126}}{m(3,3,1,-2/3)_{126}}}
+\ln{\frac{m(3,1,1,-2/3)_{126}}{m(1,2,2,0)_{126}}}
- 2\ln  {M_X\over m(8,2,2,0)_{126}} \right] \nonumber \\
&+& {1\over11}\left[ \ln{\frac{M_X}{m(1,3,1,0)_{210}}}
+\frac{9}{2}\ln{\frac{m(8,1,3,0)_{210}}{m(8,3,1,0)_{210}}}
+\frac{3}{2}\ln{\frac{m(3,1,3,4/3)_{210}}{m(3,3,1,4/3)_{210}}}
\right. \nonumber \\
&+& \frac{15}{2}\ln{\frac{m(\overline{6},1,3,-2/3)_{126}}{m(6,3,1,2/3)_{126}}}
+ \frac{3}{2}\ln{\frac{m(\overline{3},1,3,2/3)_{126}}{m(3,3,1,-2/3)_{126}}}
+ \ln{\frac{m(3,2,2,4/3)_{126}}{m(\overline{3},2,2,-4/3)_{126}}} \nonumber \\
&+& \left.
\frac{1}{2}\ln{\frac{m(\overline{3},1,1,2/3)_{126}}{m(3,1,1,-2/3)_{126}}}
+ \ln{\frac{m(3,1,1,-2/3)_{10}}{m(3,1,1,-2/3)_{126}}} \right] \nonumber \\
B &=& \frac{1}{2} \ln{\frac{M_X M_R}{M_Z^2}}
+{1\over11}\left[ \ln{\frac{m(6,2,2,2/3)_{210}}{m(1,2,2,2)_{210}}}
+ {3\over2}\ln{\frac{m(8,3,1,0)_{210}}{m(3,3,1,4/3)_{210}}}
\right. \label{eq:bstrat} \\
&-& {1\over4}\ln{M_X\over m(8,1,1,0)_{210}}
+ {3\over2}\ln{\frac{M_X}{m(1,3,1,-2)_{126}}}
+2\ln{\frac{m(8,2,2,0)_{126}}{m(3,2,2,4/3)_{126}}} \nonumber \\
&-& \left. {3\over2}\ln{\frac{M_X}{m(6,3,1,2/3)_{126}}} \right]
+ {1\over11}\left[ \frac{5}{4}\ln{\frac{M_X}{m(1,1,3,2)_{126}}}
+ \frac{5}{4}\ln{\frac{m(8,3,1,0)_{210}}{m(8,1,3,0)_{210}}} \right. \nonumber
\\
&+& \frac{9}{4}\ln{\frac{m(3,3,1,4/3)_{210}}{m(3,1,3,4/3)_{210}}}
+ \frac{1}{4}\ln{\frac{m(1,3,1,0)_{210}}{m(1,1,3,0)_{210}}}
- \frac{5}{4}\ln{\frac{M_X}{m(1,3,1,-2)_{126}}} \nonumber \\
&+& \left.
\frac{9}{4}\ln{\frac{m(6,3,1,2/3)_{126}}{m(\overline{6},1,3,-2/3)_{126}}}
+ \frac{3}{2}\ln{\frac{m(3,3,1,-2/3)_{126}}{m(\overline{3},1,3,2/3)_{126}}}
+ \ln{\frac{m(3,2,2,4/3)_{126}}{m(\overline{3},2,2,-4/3)_{126}}}
\right]. \nonumber
\enqa}

At one- and two-loops, one should get, with the ESH:
\[\bea{lll}
M^{(1)}_R=8.3\cdot10^{8}\cdot3.3^{0\pm1} & M^{(1)}_X=3.5\cdot10^{16}
\cdot2.1^{0\pm1} &\t^{(1)}=1.4\cdot10^{37}\cdot2.1^{0\pm4} \\
M^{(2)}_R=2.7\cdot10^{9} & M^{(2)}_X=1.1\cdot10^{16}
& \t^{(2)}=1.5\cdot10^{35}.
\ena\]

In a previous paper (Buccella and Rosa, 1992) one considered a particular
direction,
\bee
\gp_{ABRS}\equiv\gp_0\left(\sqrt{\frac{3}{5}},0,\sqrt{\frac{2}{5}}\right)=
\sqrt{\frac{3}{5}}\gp_L+\sqrt{\frac{2}{5}}\gp_T,
\ene
for the breaking at the highest scale and, for the model introduced in
Abud, Buccella, Rosa, and Sciarrino (1989), one has been able to deduce,
from the inequalities found for the
masses and eq. (\ref{eq:bstrat}), the inequality
\bee
M_R\leq\frac{M_{SU(5)}^2}{M_X}\leq 5\cdot 10^{11}~GeV,
\ene
the last inequality being deduced from the lower limit on $M_X$ in eq.
(\ref{eq:mxinf}).

One expects to find the absolute minimum of $V_\gp$ around the
direction $\gp_{ABRS}$ if the coefficient of the invariant
$||(\gp\gp)_{54}||$, vanishing at $\gp_{ABRS}$, is positive and larger
than the others. Indeed, it can be
shown that for $\bl\neq0$ one can get values of the parameters consistent
with $\tilde z_1=\frac{\bl+\sqrt{\bl^2-\al\cl}}{\al}=\sqrt{\frac{3}{5}}$
and with the necessary conditions described in table III.

Here, we are looking for the highest value for $M_R$ consistent with the
necessary conditions obtained in the thesis of one of us (Pisanti, 1992)
to get the minimum in the desired direction.

Due to the complexity of the conditions, we have not been able to deduce
analytically, as in the previous cases, interesting inequalities for the
one-loop equations.
It has been therefore necessary to proceed numerically
(Rosa, 1993) to the search of the highest value for $M_R$.
The following predictions for $M_R$, $M_X$, and $\t$,
which were obtained with the scalars of the
$126\oplus \overline{126}$ at the scale $M_R$ and with the scalars of the
210 at the scale $M_X$, are the result of this numerical analysis:
\[\bea{lll}
M^{(1)}_R=4.8\cdot10^{10}\cdot2.5^{0\pm1} & M^{(1)}_X=2.8\cdot10^{16}
\cdot2.0^{0\pm1} &\t^{(1)}=6.1\cdot10^{36}\cdot2.0^{0\pm4} \\
M^{(2)}_R=1.2\cdot10^{11} & M^{(2)}_X=1.9\cdot10^{16}
& \t^{(2)}=1.2\cdot10^{36}.
\ena\]

\section*{IV. Conclusions}
We have studied the predictions for the values of the scale of spontaneous
breaking of the intermediate symmetry $G'\supset SU(2)_R$ for a class of
$SO(10)$ models. In table IV we report the values of the upper limits found
for $M_R$ by demanding that, from proton stability,
$M_X\geq3.2\cdot 10^{15}~GeV$.

For the model with $G'\supset SU(4)_{PS}\times D$, which is the one giving rise
to the largest value for $M_R$, the central prediction for $M_X$ is two
standard deviations away from the experimental lower limit.

The upper limit for $M_R$ for the model with
$\frac{G'}{SU(2)_R\otimes SU(2)_L}=SU(4)_{PS}$ is almost two orders of
magnitude larger than the value found with the ESH.

\begin{table}
\centerline{TABLE IV}
{\scriptsize
\begin{center}
\begin{tabular}{|c|c|c|} \hline
          &                             &   \\
$\frac{G'}{SU(2)_R\otimes SU(2)_L}$ & $\frac{M^{(2)}_R}{10^{11}~GeV}$ &
$\frac{M^{(2)}_X}{3.2\cdot10^{15}~GeV}$ \\
          &          &        \\ \hline
          &                             &   \\
$SU(4)_{PS}\times D$ & $ 360\cdot(1.3)^{0\pm1} $ & $ 0.34\cdot(1.8)^{0\pm1}
$ \\
          &                             &   \\ \hline
          &                             &   \\
$SU(4)_{PS}$  &  $ 110\cdot(2.5)^{0\pm1} $ & $ 0.97\cdot(1.8)^{0\pm1} $ \\
          &                             &   \\ \hline
          &                             &   \\
$SU(3)_C\otimes U(1)_{B-L}\times D$ & $ 0.29\cdot(3.1)^{0\pm1} $ &
$ 0.97\cdot(1.9)^{0\pm1}  $ \\
          &                              &  \\ \hline
          &                             &   \\
$SU(3)_C\otimes U(1)_{B-L} $ & $ 1.2\cdot(2.5)^{0\pm1} $  &
$ 5.9\cdot(2.0)^{0\pm1}  $ \\
          &                              &  \\ \hline
\end{tabular}
\end{center}}
\end{table}

For the last two models we find an upper limit for
$M_R~(=1.2\cdot10^{11}\cdot(2.5)^{0\pm1}~GeV)$ which gives rise, within
the see-saw mechanism, to the following lower limits for $m_{\nu_\tau}$
and $m_{\nu_\mu}$:
\beqa
m_{\nu_{\tau}} &\geq& 11~ \frac{g_{2R}(M_R)}{f_3(M_R)}
\left(\frac{m_t}{100~GeV}\right)^2~eV  \\
m_{\nu_{\mu}}  &\geq& 2.4\cdot10^{-3}~\frac{g_{2R}(M_R)}{f_2(M_R)}~eV.
\enqa
(these formulae are obtained using the following values:
$m_\tau=1784.1~MeV,~m_c=1500~MeV$, ${m_b=5000~MeV}$ (Review of Particle
Properties, 1992);
$g_{2R}$ and $f_i$ are the $SU(2)_R$ gauge coupling constant and the
Yukawa coupling of the $126\oplus\overline{126}$ to the i-th family
respectively.)

These values imply a substantial contribution of $\nu_\tau$ to the dark
matter in the universe and a value for $\nu_\mu$ which might be relevant
for the solution of the solar-neutrino problem in terms of the MSW model.
With respect to a previous analysis by two of us (Buccella and Rosa, 1992)
the results
for the model with $G'\supset SU(4)_{PS}\times D$ are modified mainly by the
recent slight increase in the experimental value of $\al_s(M_Z)$.

For the models with $G'\supset SU(4)_{PS}$ or $SU(3)_C\otimes
U(1)_{B-L}\times D$ the region allowed for the parameters
is reduced by the necessary requirement that the absolute minimum in
the 3-dimensional
$SU(3)_c \otimes SU(2)_L \otimes U(1)_{T_{3R}} \otimes U(1)_{B-L}$-stratum
is $V(\gp_T)$ or $V(\gp_L)$ respectively.

For the model with $G'\equiv G_{3221}$ the present analysis concerns
all values of the parameters complying with the conditions which are
necessary for the desired symmetry breaking pattern, while in
Buccella and Rosa (1992) only the particular choice of
Abud, Buccella, Rosa, and Sciarrino (1989) was considered.

The conclusion of our rather general analysis of the $SO(10)$ models
is that by requiring agreement with the present lower limit for $\t$ one
finds neutrino masses for $\nu_{\tau}$ and $\nu_{\mu}$ of the order of
magnitude relevant for cosmology and solar-neutrino astrophysics.

The model with $G' \supset SU(4)_{PS}\times D$, for which smaller
$\nu_\tau$ masses are expected, is almost excluded by
experimental information on proton decay.

The model with $G' \supset SU(4)_{PS}$, which however within the ESH predicts a
value of $m_{\nu_{\mu}}=2.0\cdot 10^{-3}\frac{g_{2R}(M_R)}{f_2(M_R)}~eV$,
may predict lower values ($\sim 10^{-5}\frac{g_{2R}(M_R)}{f_2(M_R)}~eV$)
only if ad-hoc assumptions for the
masses of the $SU(2)_{L(R)}$ triplets of the 126 are made, while the
contribution to RGE of the triplets of the 210 is controlled by the
necessary condition in eq. (\ref{eq:mastiz}).

A better knowledge of the gauge coupling constants at $M_Z$, especially of
$\al_S$, as well as the increase of the lower limit on $\t$ would improve
the predictive power of the analysis here described.

\section*{Appendix}

\bigskip

\centerline{210}

\bigskip
{
\beqsa
\frac{m^2(8,3,1,0)}{r^2} &=&
\al\left(\frac{3}{4}-\frac{3}{4}z_1^2-\frac{1}{2}z_1^4
+\frac{\sqrt{3}}{2} z_1z_3\right)+
\bl\left(z_1+z_1^3+\sqrt{3} z_3\right) \\
 & & +\cl\left(\frac{3}{2}-\frac{5}{2}z_1^2+\frac{7}{2 \sqrt{3}}z_1z_3\right)+
\dl\left(-\frac{40}{9}z_1^2+\frac{40}{3 \sqrt{3}}z_1z_3\right)   \\  &  &  \\
\frac{m^2(8,1,3,0)}{r^2} &=&
\al\left(\frac{3}{4}-\frac{3}{4}z_1^2-\frac{1}{2}z_1^4
-\frac{\sqrt{3}}{2} z_1z_3\right)+
\bl\left(z_1+z_1^3-\sqrt{3} z_3\right) \\
& & +\cl\left(\frac{3}{2}-\frac{5}{2}z_1^2-\frac{7}{2 \sqrt{3}}z_1z_3\right)-
\dl\left(\frac{40}{9}z_1^2+\frac{40}{3 \sqrt{3}}z_1z_3\right)   \\  &  &  \\
\frac{m^2(3,3,1,4/3)}{r^2} &=&
\al\left(\frac{3}{4}-\frac{1}{4}z_1^2-\frac{1}{2}z_1^4
-\frac{\sqrt{3}}{2}z_1z_3\right)+\bl\left(-z_1+z_1^3+\sqrt{3}z_3\right) \\
& & +\cl\left(\frac{3}{2}-\frac{3}{2}z_1^2-\frac{7}{2\sqrt{3}} z_1z_3 \right)
-\dl\left(\frac{20}{9}z_1^2+\frac{40}{3\sqrt{3}} z_1z_3 \right)  \\  &  &  \\
\frac{m^2(3,1,3,4/3)}{r^2} &=&
\al\left(\frac{3}{4}-\frac{1}{4}z_1^2-\frac{1}{2}z_1^4
+\frac{\sqrt{3}}{2}z_1z_3\right)+\bl\left(-z_1+z_1^3-\sqrt{3}z_3\right) \\
& & +\cl\left(\frac{3}{2}-\frac{3}{2}z_1^2+\frac{7}{2\sqrt{3}} z_1
z_3 \right)-\dl\left(\frac{20}{9}z_1^2-\frac{40}{3\sqrt{3}} z_1z_3
\right)  \\  &  &  \\
\frac{m^2(1,3,1,0)}{r^2} &=&
\al\left(\frac{3}{4}+\frac{3}{4}z_1^2-\frac{1}{2}z_1^4
-\sqrt{3}z_1z_3\right)+\bl\left(-2z_1+z_1^3+\sqrt{3}z_3\right) \\
& & +\cl\left(\frac{3}{2}+z_1^2-\frac{7}{\sqrt{3}}z_1z_3\right)+
\dl\left(\frac{80}{9}z_1^2-\frac{80}{3\sqrt{3}}z_1z_3\right)  \\  &  &  \\
\frac{m^2(1,1,3,0)}{r^2} &=&
\al\left(\frac{3}{4}+\frac{3}{4}z_1^2-\frac{1}{2}z_1^4
+\sqrt{3}z_1z_3\right)+\bl\left(-2z_1+z_1^3-\sqrt{3}z_3\right) \\
& & +\cl\left(\frac{3}{2}+z_1^2+\frac{7}{\sqrt{3}}z_1z_3\right)+
\dl\left(\frac{80}{9}z_1^2+\frac{80}{3\sqrt{3}}z_1z_3\right) \\  &  &  \\
\frac{m^2(8,1,1,0)}{r^2} &=&
-\frac{1}{2}\al z_1^4+\bl\left(z_1+z_1^3\right)+\frac{1}{2}\cl
\left(1-3z_1^2\right) \\  &  &  \\
\frac{m^2(3,1,1,4/3)}{r^2} &=&
\frac{1}{2}\al\left(z_1^2-z_1^4\right)+\bl\left(-z_1+
z_1^3\right)+\frac{1}{2}\cl\left(1-z_1^2\right)  \\  &  &  \\
\frac{m^2(6,2,2,2/3)}{r^2} &=&
-\frac{1}{2}\al z_1^4+\bl\left(z_1+z_1^3\right)-\cl z_1^2+
\dl\left(-5+\frac{5}{3}z_1^2\right)  \\  &  &  \\
\frac{m^2(1,2,2,2)}{r^2} &=&
\al\left(3z_1^2-\frac{1}{2}z_1^4\right)+\bl\left(-3z_1+z_1^3\right)+\cl 5
z_1^2+\dl\left(-5+15z_1^2\right) \\  &  &  \\
\frac{m^2(3,2,2,-2/3)_{diag}}{r^2} &=&
{}~~\left\{\bea{ll}
\frac{\left(\cl+10\dl\right)\left(-3+z_1^2\right)}{6} &
SU(3)_c\otimes SU(2)_L\otimes SU(2)_R\otimes U(1)_{B-L} \\ & \\
-5\dl & SU(4)_{PS}\otimes SU(2)_L\otimes SU(2)_R \\ & \\
\frac{\left(-\al+2\bl-\frac{5}{3}\cl-\frac{20}{3}\dl\right)}{2}
& SU(3)_c\otimes SU(2)_L\otimes SU(2)_R\otimes U(1)_{B-L} \times D
 \ena \right.
\enqsa}

\bigskip

\centerline{ $126\oplus\overline{126}$}

{
\beqsa
\frac{m^2(6,3,1,2/3)}{s^2}  &=&
f_4 \frac{1}{75}\left(2 z_1+\sqrt{3}z_3\right)^2 \\
 & & \\
\frac{m^2(3,3,1,-2/3)}{s^2} &=&
f_2 \frac{4}{315} z_1^2 +f_3 \frac{8}{315}z_1^2 +f_4
\frac{1}{75} \left(z_1+\sqrt{3} z_3\right)^2  \\     & & \\
\frac{m^2(1,3,1,-2)}{s^2} &=&
f_4 \frac{1}{25}z_3^2 \\  & & \\
\frac{m^2(8,2,2,0)}{s^2} &=&
f_2 \frac{1}{105} z_3^2 +f_3\frac{2}{105}\left(z_1+z_3\right)^2 +
f_4 \frac{1}{300}\left(4 z_1+\sqrt{3} z_3\right)^2  \\  & & \\
\frac{m^2(3,2,2,4/3)}{s^2} &=&
f_2 \frac{1}{105}\left(z_3+\sqrt{\frac{4}{3}} z_1\right)^2 +
f_3 \frac{2}{315}\left(z_1-\sqrt{3} z_3\right)^2 +
f_4\frac{1}{300} \left(2  z_1+\sqrt{3} z_3\right)^2 \\   & & \\
\frac{m^2(\bar3,2,2,-4/3)}{s^2} &=&
f_2 \frac{1}{105}\left(z_3-\sqrt{\frac{4}{3}}z_1\right)^2
+ f_3 \frac{2}{315} \left(z_1+\sqrt{3} z_3\right)^2 +
 f_4 \frac{1}{300}\left(2  z_1+\sqrt{3} z_3\right)^2 \\   & & \\
\frac{m^2(1,2,2,0)}{s^2} &=&
f_1 \frac{1}{126}z_1^2 +f_2 \frac{1}{105}z_3^2 +f_3
\frac{2}{105}\left(z_3+\sqrt{\frac{4}{3}} z_1\right)^2 +
f_4 \frac{1}{300}\left(z_1+\sqrt{3} z_3\right)^2 \\  & & \\
\frac{m^2(\bar6,1,3,-2/3)}{s^2} &=&
f_4 \frac{4}{75} z_1^2 \\  & & \\
\frac{m^2(\bar3,1,3,2/3)}{s^2} &=&
f_2 \frac{4}{315}z_1^2 +f_3 \frac{8}{315}z_1^2+
f_4 \frac{1}{75}z_1^2 \\   & & \\
\frac{m^2(3,1,1,-2/3)}{s^2} &=&
f_1 \frac{1}{378}\left(z_1-\sqrt{3} z_3\right)^2 +
f_2 \frac{2}{315}z_1^2 + f_3 \frac{4}{63}\left(z_3 +
\frac{4}{\sqrt{15}} z_1 \right)^2 + f_4
\frac{1}{100}\left(z_3+\sqrt{3} z_1\right)^2 \\  & & \\
\frac{m^2(\bar3,1,1,2/3)}{s^2} &=&
f_1 \frac{1}{378}\left(z_1+\sqrt{3} z_3\right)^2 +
f_2 \frac{2}{315} z_1^2 + f_3 \frac{4}{63}\left(z_3-
\frac{4}{\sqrt{15}} z_1 \right)^2 + f_4 \frac{1}{100}\left(z_3+\sqrt{3}
z_1\right)^2
\enqsa }

\section*{References}

Abud, M., F. Buccella , A. Della Selva, A. Sciarrino, R. Fiore,
and G. Immirzi, 1986, Nucl.
\indent
Phys. B {\bf 263}, 336.

\noindent
Abud, M., F. Buccella, L. Rosa, A. Sciarrino, 1989,
Zeitschrift $f\ddot u$r Phys. C {\bf 44}, 589.

\noindent
Amaldi, U., W. de Boer, and H. Furstenau, 1991, Phys. Lett. B {\bf 260}, 447.

\noindent
Amelino-Camelia, G., 1989, {\it``Spontaneous Symmetry Breaking in
SO(10) GUT Models with
\indent $\tau_p\sim 10^{33}$ years"} First-Degree Thesis
(unpublished).

\noindent
Amelino-Camelia, G., F. Buccella, and L. Rosa, 1990, Atti del
{\it``Second International \linebreak \indent
Workshop on Neutrino Telescopes"} ed. Milla
Baldo Ceolin.

\noindent
Anselmo, F., L. Cifarelli, and A. Zichichi, 1992, Il Nuovo Cimento A
{\bf 105} n.9, 1357.

\noindent
Appelquist,T., and J. Carazzone, 1975, Phys. Rev. D {\bf 11} n.10, 2856.

\noindent
Babu, K. S., and Q. Shafi, 1993, Nucl. Phys. B (Proc. Suppl.) {\bf 31}, 242.

\noindent
Barbieri, R., G. Morchio, D.V. Nanopoulos, and F. Strocchi, 1980, Phys.
Lett. B {\bf 90}, 91.

\noindent
Baseq, J., S. Meljanac, and L. O'Raifeartaigh, 1989, Phys. Rev. D
{\bf 39}, 3110.

\noindent
Buccella, F., 1988, in {\it``Symmetry in Science $III$"}, eds.
B. Gruber. and F. Iachello.

\noindent
Buccella, F., L. Cocco, A. Sciarrino, and T. Tuzi, 1986, Nucl. Phys.
B {\bf 274}, 559.

\noindent
Buccella, F., L. Cocco, and C. Wetterich, 1984, Nucl. Phys. B {\bf 243}, 273.

\noindent
Buccella, F., G. Miele, L. Rosa, P. Santorelli, and T. Tuzi, 1989,
Phys. Lett. B {\bf 233}, 178.

\noindent
Buccella, F., and L. Rosa, 1987, Zeitschrift f$\ddot u$r Phys. C {\bf 36}, 425.

\noindent
Buccella, F., and L. Rosa, 1992, Nucl. Phys. B (Proc. Suppl.) {\bf 28 A}, 168.

\noindent
Buccella, F., and H. Ruegg, 1982, Il Nuovo Cimento A {\bf 67} n.1, 61.

\noindent
Burzlaff, J., and L. O'Raifeartaigh, 1990, Proceedings of Capri Symposia,
1983-1987, ed. F.
\indent
Buccella.

\noindent
Chang, D., J. M. Gipson, R. E. Marshak, R. N. Mohapatra, and M. K. Parida,
1985, Phys.
\indent
Rev. D {\bf 31}, 1718.

\noindent
Chang, D., R. N. Mohapatra, and M. K. Parida, 1984, Phys. Rev.
Lett. {\bf 52}, 1072.

\noindent
Deshpande, N. G., E. Keith, and Palash B. Pal, 1992, Phys. Rev. D
{\bf 46} n.5, 2261.

\noindent
Dixit, V. V., and M. Sher, 1989, Phys. Rev. D {\bf 40} n.11, 3765.

\noindent
Fritzsch, H., and P. Minkowski, 1975, Ann. of Phys. {\bf 93}, 183.

\noindent
Gell-Mann, M., P. Ramond, and R. Slansky, 1980, {\it``Supergravity"}
(North Holland, \linebreak \indent Amsterdam).

\noindent
Georgi, H., 1975, {\it``Particles and Fields"}  C. E. Carlson (AJP).

\noindent
Jones, D. R. T., 1982, Phys. Rev. D {\bf 25} n.2, 581.

\noindent
Kaymackcalan, O., L. Michel, K.C. Wali, W.D. Mc GLinn, and L.
O'Raifeartaigh, 1986,
\indent
Nucl. Phys. B {\bf 267} n.1, 203.

\noindent
Kuzmin, V., and N. Shaposhnikov, 1980, Phys. Lett. B {\bf 92}, 115.

\noindent
Mohapatra, R. N., and M.K. Parida, 1993, Phys. Rev. D {\bf 47}, 264.

\noindent
Pati, J. C., and A. Salam, 1974, Phys. Rev. D {\bf 10}, 275.

\noindent
Pisanti, O., 1992, {\it``An Analysis of SO(10) Models with
Intermediate Symmetry $SU(3)_c\otimes
\indent
SU(2)_L \otimes SU(2)_R \otimes
U(1)_{B-L}$"} First-Degree Thesis (unpublished).

\noindent
Review of Particle Properties, 1992, Phys. Rev. D {\bf 45} n. 11.

\noindent
Rosa, L., 1993, {\it``Neutrino Masses and Proton Lifetime in SO(10)
GUT Models"} Ph. D.
\indent Thesis (unpublished).

\noindent
Schramm, D., 1992, Nucl. Phys. B (Proc. Suppl.) {\bf 28 A}, 243.

\noindent
Shafi, Q., and C. Wetterich, 1979, Phys. Lett. B {\bf 85}, 52.

\noindent
Tuzi, T., 1989, {\it``Two-loop Contributions to Spontaneous Breaking
Scales in SO(10)"} Ph. D.
\indent
Thesis (unpublished).

\noindent
Weinberg, S., 1980, Phys. Lett. B {\bf 91}, 51.

\noindent
Yanagida, T., 1979, {\it Proceedings of the Workshop on the Unified Theory
and the Baryon
\indent
Number of the Universe} edited by O. Sawada {\it et al.} (KEK).

\noindent
le Yaouanc, A., L. Oliver, O. P\`ene, and J.C. Raynal, 1977, Phys. Lett.
B {\bf 72}, 53.

\end{document}